\documentclass[referee]{raa}            

\usepackage{graphicx,times}             
\usepackage{natbib}
\usepackage{amssymb,amsmath}
\bibpunct{(}{)}{;}{a}{}{,}

\usepackage[pagebackref=true]{hyperref}
\usepackage{graphicx}	
\usepackage{amsmath}	
\usepackage{amssymb}	

\newcommand{\kms}{\rm{km\,s^{-1}}} 
\newcommand{\kpc}{\rm{kpc}}

\newcommand{\kmss}{\rm{km^2\,s^{-2}}}
\newcommand{\masyr}{\rm{mas\,yr^{-1}}}
\newcommand{\dplx}{d_{\varpi}}
\newcommand{\rgeo}{r_{\rm{geo}}}
\newcommand{\vsun}{v_\odot}
\newcommand{\Usun}{U_\odot}
\newcommand{\Vref}{V_{\rm{ref}}}
\newcommand{\vl}{v_l}
\newcommand{\vlref}{v_{l,\rm{ref}}}
\newcommand{\dvl}{\Delta v_l}
\newcommand{\ikpc}{\rm{kpc}^{-1}}
\newcommand{\sqrtP}{\sqrt{\rm{P}}}
\newcommand{\sqrtPlf}{\sqrt{\rm{P_{LF}}}}

\begin{document}

  \title{Power spectrum of longitudinal-velocity residuals in the Galactic disk from Gaia DR3 red-giant-branch stars
}

   \volnopage{Vol.0 (202x) No.0, 000--000}      
   \setcounter{page}{1}          

   \author{Ping-Jie Ding 
      \inst{}
   }

   \institute{XinJiang Astronomical Observatory, Chinese Academy of Sciences,150 Science 1-Street, Urumqi, Xinjiang 830011, China; {\it dingpingjie@xao.ac.cn}\\
\vs\no
   {\small Received 202x month day; accepted 202x month day}}

\abstract{Non-axisymmetric structures in the Galaxy produce coherent stellar motions in the Galactic disk. These motions are usually interpreted by assuming parameterized gravitational potentials or specific perturbation models, thereby entangling direct kinematic information with model-dependent assumptions. Using Gaia DR3 astrometry, we analyze the Galactic longitudinal-velocity residuals of nearly four million red-giant-branch stars over $-4<X_h<4\,\kpc$ and $-6<Y_h<6\,\kpc$ in the heliocentric region. We construct an empirical axisymmetric baseline without introducing any dynamical model, and compute the power spectrum of the residual velocity field. The spectrum exhibits a broad peak at spatial frequency of $f\simeq0.12–0.22\,\ikpc$, corresponding to a scale of around $5-8\,\kpc$. The residual amplitude is sensitive to the adopted distance scale: inverse-parallax distances $\dplx$ yield a peak of around $16\,\kms$, while the Bailer-Jones geometric distances $\rgeo$ yield around $12\,\kms$. Minimizing the low-frequency power favors a solar radial motion of $\Usun=10.8\,\kms$ based on $\dplx$ and $\Usun=10.4\,\kms$ based on $\rgeo$. Collectively, the resulting spectrum establishes Gaia astrometry as a direct probe of disk-wide non-axisymmetric kinematics and provides quantitative targets for future dynamical modeling and interpretation. 
}

   \authorrunning{P.-J. Ding}            
   \titlerunning{Power spectrum of $\vl$ residuals}  

   \maketitle
\section{Introduction}\label{sect1}

The Galactic disk is not axisymmetric. Perturbers including the central bar (\citealt{Bovy19}; \citealt{Bernet24}; \citealt{YB26}), a potentially triaxial halo (\citealt{Woudenberg24}; \citealt{Dillamore26}; \citealt{Zhu26}), spiral arms (\citealt{Hou21}; \citealt{Poggio21}; \citealt{Lin26}) and tidal streams (\citealt{Jerabkova21}; \citealt{Zhou26}) can produce deviations from circular rotation. Gaia observations have revealed the presence of non-axisymmetric patterns in stellar kinematics across a substantial fraction of the disk (e.g., \citealt{Eilers20}; \citealt{McMillan22}; \citealt{Gaia23b}; \citealt{Fedorov26}). However, interpretation of these structures usually involves assumptions of the gravitational potential, perturbation model, distribution function, etc. While such dynamical models are essential for uncovering physical parameters, they naturally introduce model dependencies and degeneracies, rendering conclusions sensitive to the specific choices of priors and assumptions.

An alternative is to first describe the observed velocity field and perform a statistical analysis to identify the spatial scales of non-axisymmetric structures, without assigning it to a specific dynamical mechanism. Power spectrum (PS) estimation is particularly well suited for this purpose, as it quantifies the amplitude of velocity fluctuations as a function of spatial frequency. This approach has been applied by \cite{Bovy15} to analyze the line-of-sight velocity field of stars from the APOGEE (primary sample, containing approximately 8000 red clump stars with a typical line-of-sight velocity uncertainty of $0.1\,\kms$), RAVE, and GCS surveys, revealing significant velocity power on scales of $\lesssim5\,\kpc$. Their analysis, however, was limited by the reliance on a prior dynamical model serving as an axisymmetric baseline and the data available at the time. 

The Gaia mission has greatly revolutionized the observational capabilities: Gaia DR3 \citep{Gaia23a} provides full five-parameter astrometric solutions for approximately 1.47 billion sources, with median parallax and annual proper motion uncertainties of 0.02–0.03 mas at magnitude $G = 9$–14 and $\sim0.5$ mas at $G=$ 20 \cite{Lindegren21a}. In this study, we perform a comprehensive power spectral analysis of nearly four million red-giant-branch (RGB) stars with high-quality astrometric measurements from Gaia DR3. We employ Galactic longitude velocity $v_l$, an ideal kinematic tracer of the Galactic disk that can be derived without spectroscopic measurements of line-of-sight velocities, as observable. In order to obtain the residual velocity field for studying the non-axisymmetric signals, we construct an empirical reference field, namely a velocity field independent of Galactocentric azimuth, directly from observables, requiring only the geometric circular-motion projection, the Galactocentric distance of the Sun, and the solar velocity relative to the Galactic center (GC). The residual velocity then measures the deviation between the observation and the reference. 

The azimuthal component of solar velocity relative to the GC is derived from the longitudinal proper motion of Sgr A*, denoted $\mu_{l,\,\rm{SgrA^*}}$. Throughout this work, $\mu_{l,\,\rm{SgrA^*}}$ is fixed as $-6.396\,\masyr$, which was determined by \citet{Gordon23} based on VLBI astrometry.

The goal of this work is to determine the spatial scales at which the disk velocity field departs from axisymmetry, and to evaluate whether large-scale residual power provides a global constraint on the radial component of solar velocity. The influence of distance scales on residual power is also assessed. Section \ref{sect2} introduces the sample used in this work. In Section ~\ref{sect3}, we construct the residual velocity field and perform its spectral analysis. Section ~\ref{sect4} presents the resulting PS and the estimation of solar velocity. We provide our discussion in Section \ref{sect5} and draw our conclusion in Section ~\ref{sect6}.

\section{Sample selection}\label{sect2}

\begin{figure}
   \centering
   \includegraphics[width=10cm]{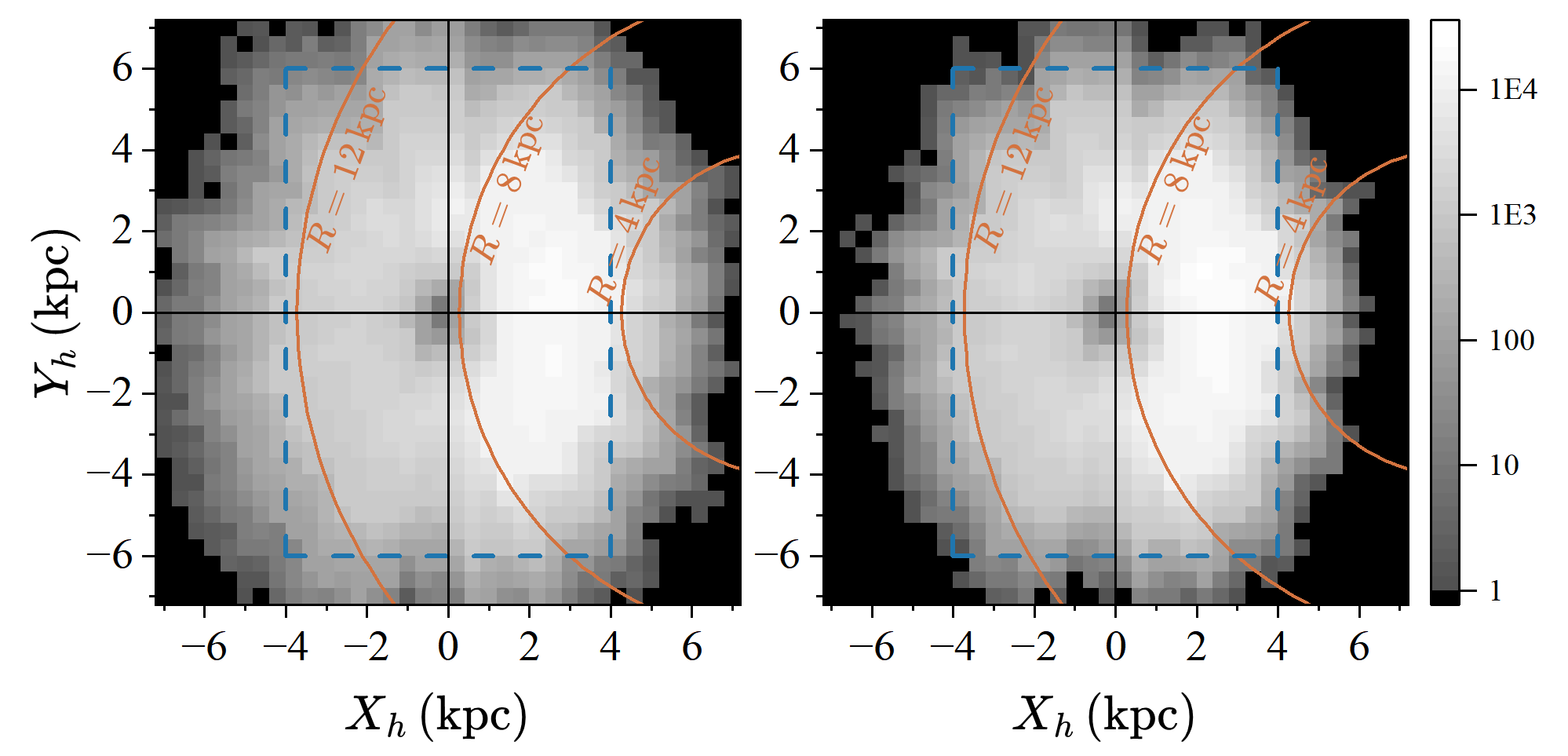}
   \caption{Number density distribution of the selected RGB stars in the heliocentric Cartesian coordinate system. The left panel shows the distribution derived from  $\dplx$; the right panel is for the distribution derived from $\rgeo$. The grayscale indicates the stellar count per $0.4\,\kpc\times0.4\,\kpc$ pixel. The concentric circular arcs correspond to the Galactocentric distances of 12 kpc, 8 kpc, and 4 kpc, respectively, assuming a solar Galactocentric distance of $R_0=8.277\,\kpc$. The blue dashed rectangle delineates the spatial selection boundary applied to the sample.}
   \label{fig_num}
\end{figure}

The sample used in this work is restricted to Gaia stars exhibiting high-quality astrometry. We select astrometrically well-behaved stars employing the following criteria: (i) goodness-of-fit statistic of the astrometric solution for the source in the along-scan direction, $\rm{ASTROMETRIC_{-}GOF_{-}AL}<3.0$; (ii) significance of excess noise, $\rm{ASTROMETRIC_{-}EXCESS_{-}NOISE_{-}SIG}\leqslant2.0$; and (iii) renormalised unit weight error, $\rm{RUWE} <1.4$ \footnote{The first two criteria are adopted following the datamodel description in the Gaia Data Release 3, Documentation release 1.3; the third criterion is a widely used empirical criterion for reducing contamination from sources with potentially problematic astrometric solutions.}. Additionally, only stars with positive parallax ($\varpi>0$) and relative parallax uncertainty $\sigma_\varpi/\varpi<0.2$ are retained. 

Stars identified as variable or with metallically $\rm{[Fe/H]\leqslant-1}$ are excluded; the latter criterion is applied to remove probable halo stars. RGB stars are selected using the effective temperature $T_{\rm{eff}}$, surface gravity $\mathrm{log}\,g$, extinction-corrected color $(G_{\rm{BP}}-G_{\rm{RP}})_0$, and absolute magnitude $M_G$, such that $T_{\rm{eff}}<5900\,\rm{K}$, $\mathrm{log}\,g<3.6$, $0.9<(G_{\rm{BP}}-G_{\rm{RP}})_0<2.2$, and $-2<M_G<2.5$ \footnote{The $T_{\rm{eff}}$ and $\mathrm{log}\,g$ criteria follow those adopted by \citet{Gaia23c}; the $(G_{\rm{BP}}-G_{\rm{RP}})_0$ and $M_G$ criteria are employed to exclude young giants or outliers. The RGB selection is broad since our analysis prioritizes a large, spatially extended, and relatively well-relaxed disk population, rather than a strictly homogeneous evolutionary sample.}. We restrict the sample to the region near the Galactic mid-plane using Galactic latitude $|b|<20^\circ$ and vertical distance $|Z|<0.5\,\kpc$ \footnote{The vertical distance $Z$ to the Galactic mid-plane is computed by accounting for the vertical offset of the Sun above the Galactic mid-plane, $Z_\odot=20.8\,\rm{pc}$ \citet{BB19}}.

Two distance scales are employed independently to generate the velocity field:
\begin{itemize}
    \item the inverse of the zero-point-corrected parallaxes, i.e., $\dplx=1/(\varpi+0.017\,\rm{mas})$, adopting a global zero-point offset $-0.017\,\rm{mas}$ \citep{Lindegren21b} in parallaxes;
    \item the geometric distance $\rgeo$ of \citet{Bailer-Jones21}.
\end{itemize}
The inverse-parallax estimate is more direct but remains affected by additional parallax systematics and nonlinear inversion; the geometric distance is more stabilized but relies on a direction-dependent prior. In the following analysis, we use the difference in the results based on the two distance estimates to assess the sensitivity to the distance scale, rather than treating either as absolute.

We restrict our sample to $-300<v_l<150\,\kms$ to get rid of velocity outliers ($0.05\%$ of the selected stars for $\dplx$; $0.02\%$ for $\rgeo$), and then remove $v_l$ measurements with random uncertainties $>30\,\kms$ ($1.7\%$ for $\dplx$; $0.9\%$ for $\rgeo$). Finally, $\sim3.93$ million stars remain in the $\dplx$ sample, and $\sim3.96$ million stars remain in the $\rgeo$ sample. 

\begin{figure}
    \centering
    \includegraphics[width=0.8\linewidth]{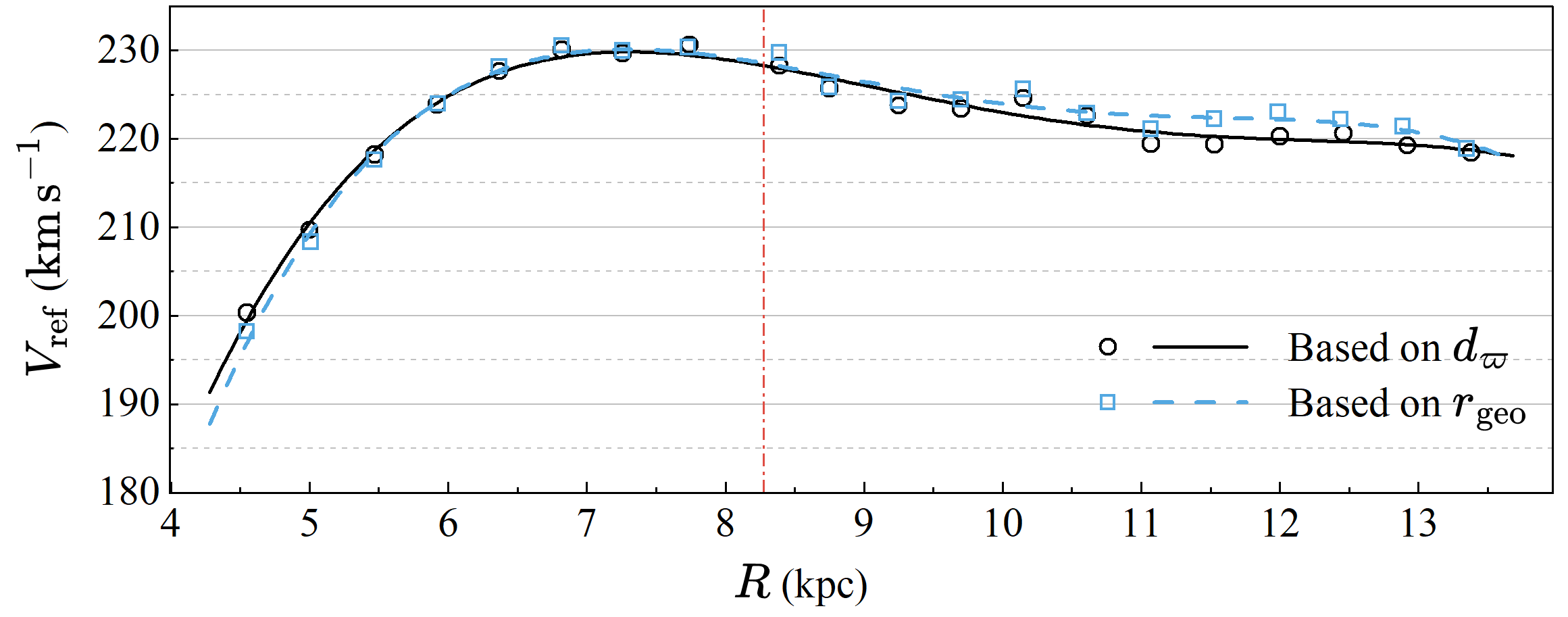}
    \caption{Axisymmetric reference baseline obtained from the RGB stars, under the assumption of $R_0=8.277\,\kpc$ and $\Usun=10\,\kms$. The black circles (based on $\dplx$) and blue squares (based on $\rgeo$) denote the medians of the axisymmetric-rotation velocity estimate $\Vref$ within each radial bin. The corresponding fourth-order polynomial fits are shown as a black solid line (for $\dplx$) and a blue dashed line (for $\rgeo$), respectively.}
    \label{fig_Vref}
\end{figure}

\begin{figure}
    \centering
    \includegraphics[width=1\linewidth]{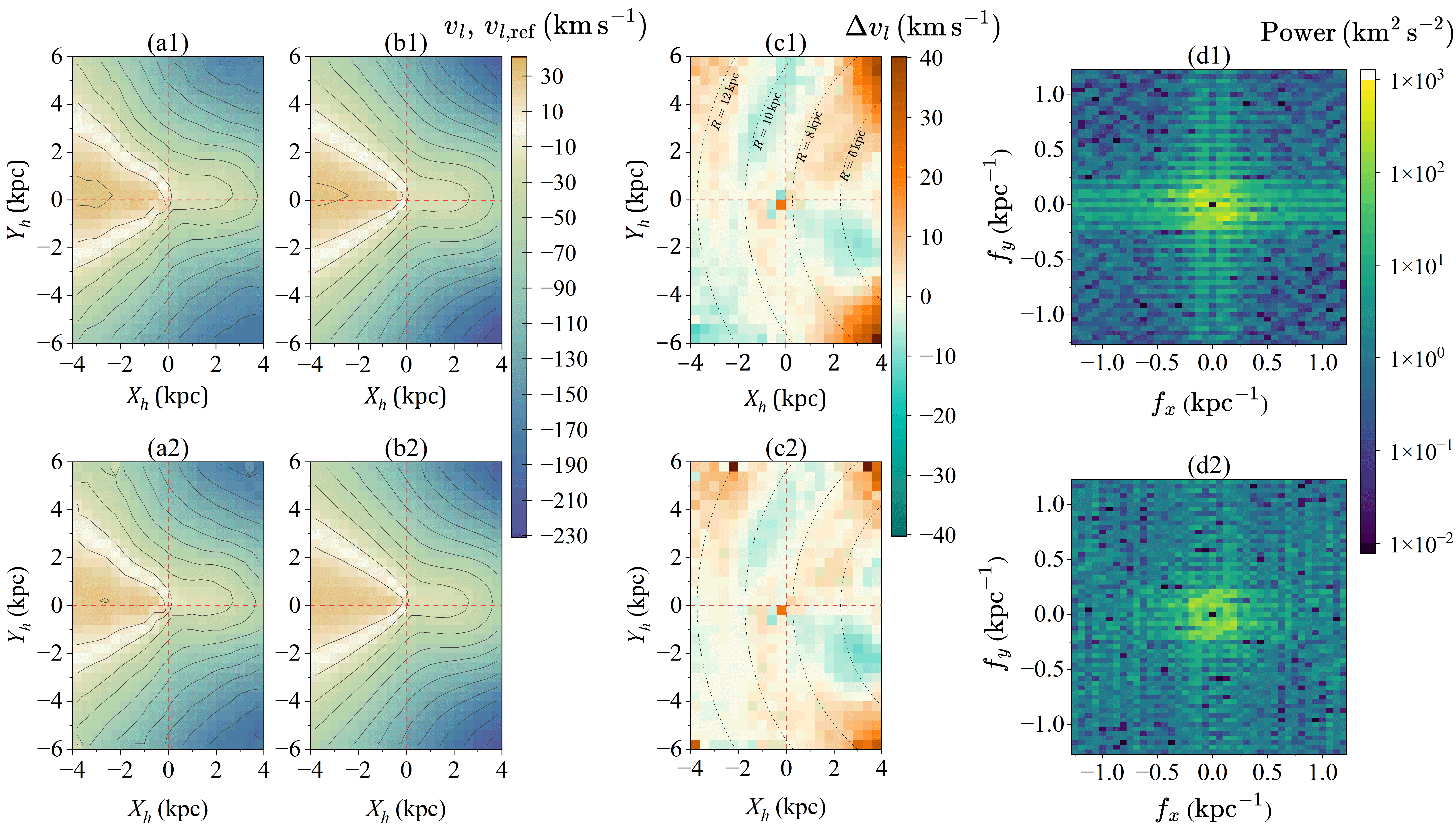}
    \caption{2-D spectral analysis of the Galactic longitude residual velocity ($\dvl$) field. Panels (a1) and (a2) display the observed $v_l$ distribution in the heliocentric Cartesian coordinate system, where color encodes the median of $v_l$ within each pixel. Panels (b1) and (b2) show the longitude reference velocity ($\vlref$) field derived from the axisymmetric reference baseline $\Vref (R)$ given by Fig. ~\ref{fig_Vref}, with color representing the median $\vlref$ per pixel. Panels (a1), (a2), (b1), and (b2) share an identical color scale. Panels (c1) and (c2) depict the residual velocity fields—defined as $\dvl\equiv v_l-\vlref$ for each star; color here denotes the median $\dvl$ per pixel; the black dotted arcs correspond to the Galactocentric distances of 12 kpc, 10 kpc, 8 kpc, and 6 kpc, respectively. Panels (d1) and (d2) present the 2-D PS of the residual field shown in (c1) and (c2), respectively; color indicates power in units of $\kmss$. The top row—panels (a1), (b1), (c1), and (d1)—corresponds to the analysis based on $\dplx$, whereas the bottom row—panels (a2), (b2), (c2), and (d2)—uses $\rgeo$.}
    \label{fig_PS2D}
\end{figure}

\begin{figure}
    \centering
    \includegraphics[width=1\linewidth]{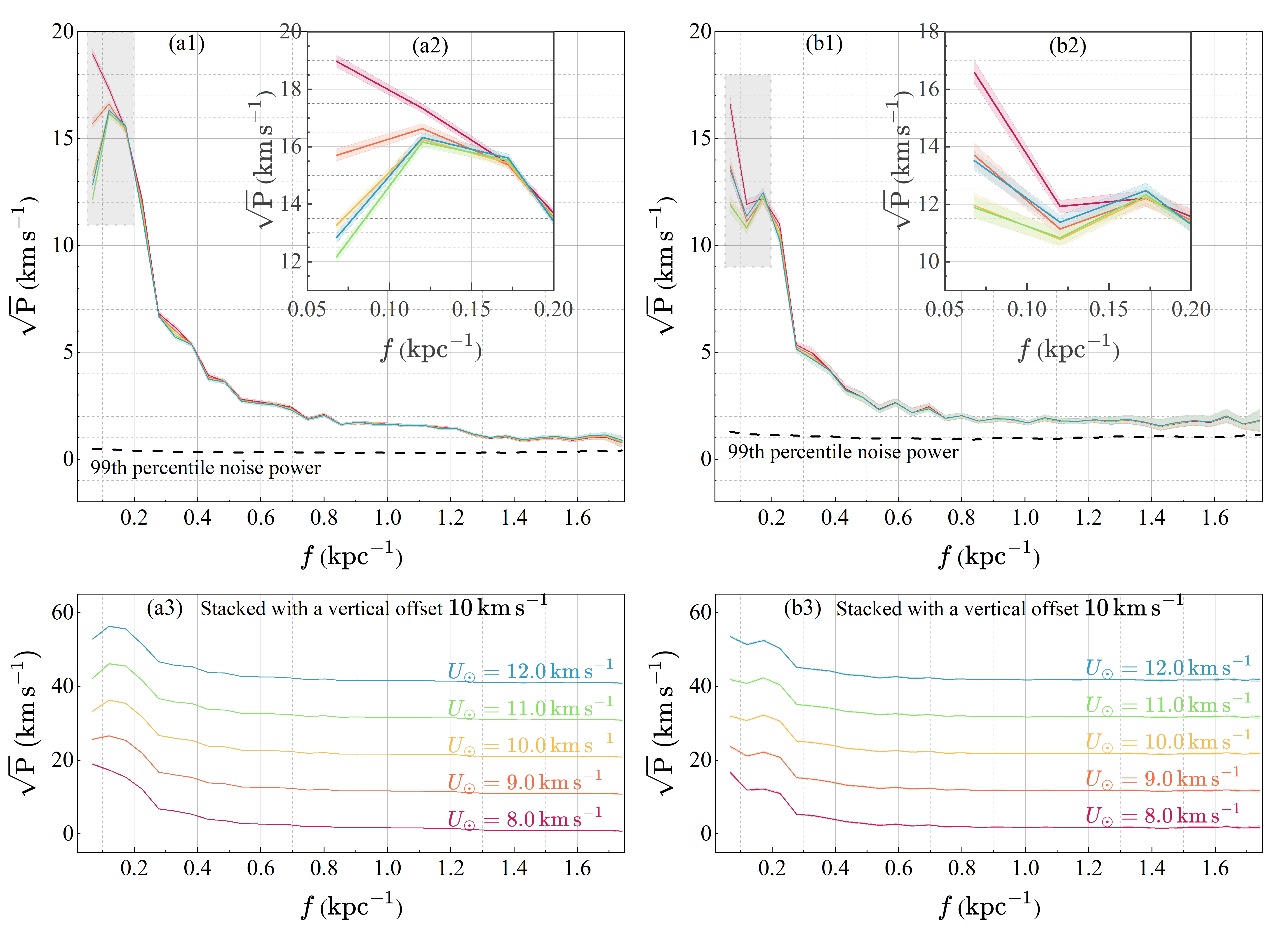}
    \caption{Amplitude $\sqrtP$ of the residual velocity field as a function of the 1-D frequency $f$. Each spectrum presented in panels (a1) and (b1) is computed based on a reference baseline constructed with a specific $\Usun$ value; the adopted $\Usun$ is encoded by color shown in panels (a3) and (b3). For each spectrum, the solid curve represents the median $\sqrtP$ obtained from mock realizations, while the surrounding shaded band indicates the interval spanning the 16th-84th percentiles (the interpercentile range can be narrow enough that the shading appears comparable in width to the line thickness). The black dashed lines mark the 99th-percentile thresholds of the noise amplitude. Panels (a2) and (b2) present zoomed views of $f<0.2\,\ikpc$ -- highlighted by light-gray shading -- in panels (a1) and (b1), respectively. Panels (a3) and (b3) show the stacked spectra corresponding to those displayed in panels (a1) and (b1), respectively. Panels (a1)-(a3) are derived using $\dplx$, whereas panels (b1)-(b3) use $\rgeo$.}
    \label{fig_PS1D}
\end{figure}

\section{Power spectral analysis of the residual velocity field}\label{sect3}

\subsection{Empirical non-axisymmetric reference and the residual field}\label{sect3.1}

To optimize the sensitivity to spatial variations associated with non-axisymmetric kinematic structures, we analyze the velocity field in a rectangular region 
\begin{equation}\label{eq_XY}
    -4<X_h<4\,\kpc,\;-6<Y_h<6\,\kpc,
\end{equation}
where $(X_h,\,Y_h)$ is the heliocentric Cartesian coordinate system, with $X_h$-axis pointing toward the GC and the $Y_h$-axis aligning with the direction of Galactic rotation. Figure \ref{fig_num} presents the stellar number density in the $(X_h,\,Y_h)$ plane. Adopting $R_0=8.277\,\kpc$ \citep{Gravity22} as the Sun's distance to the GC, the footprint spans a radial range of approximately $4.3<R< 13.7\,\kpc$.

For an axisymmetric velocity field, the predicted longitude velocity can be written as
\begin{equation}\label{eq_vlref}
    \vlref= \Vref \cos{(l+\phi)} - \vsun \cos{l} + \Usun \sin{l},
\end{equation}
where the axisymmetric reference velocity $\Vref$ is the circular-rotation velocity. Parameters $\vsun=250.96\,\kms$ (from $\mu_{l,\,\rm{SgrA^*}}$ and $R_0$) and $\Usun$ represent the Sun's azimuthal ($\phi$-, aligned with Galactic rotation) and radial (toward the GC) velocity components relative to the GC, respectively. Based on Equation~\eqref{eq_vlref}, the estimate of $\Vref$ for a star is
\begin{equation}\label{eq_Vref}
    \Vref=\frac{v_l - \Usun \sin{l} + \vsun \cos{l}}{\cos{(l + \phi)}}.
\end{equation}
For a given $\Usun$, we infer $\Vref$ for each star located within $|\phi|<10^\circ$ over $4.3<R< 13.7\,\kpc$. To avoid the numerical instability caused by a vanishing projection factor, stars used to construct $\Vref$ are restricted to $|\cos{(l + \phi)}|>0.1$. These stars are divided into twenty equal-width radial bins, and the median $R$ and $\Vref$ are obtained for each bin.

Fitting a fourth-order polynomial of $(R-R_0)$ to the data, we obtain the mean rotation of the RGB sample under a circular projection, $\Vref (R)$, which is used as a reference baseline. Note that $\Vref (R)$ is not a measurement of the gravitational circular speed and should not be interpreted as a conventional rotation curve; it is only used as an empirical baseline that removes the smooth axisymmetric part of stellar motion. All following interpretations are based on the residual field. 

The choice of $\Usun$ is drawn from literature. Depending on the tracers, estimates of $\Usun$ typically range from $8\,\kms$ to $12\,\kms$ (e.g., \citealt{Zhu09}; \citealt{Schonrich10}; \citealt{Bovy12}; \citealt{Golubov13}; \citealt{Tian15}; \citealt{Ding19}; \citealt{Reid19}; \citealt{BB23}; \citealt{ZP25}). In this case, we first adopt $\Usun=8,\,9,\,10,\,11,\,12\,\kms$, and the complete construction is performed for each assumed $\Usun$. Figure ~\ref{eq_Vref} presents the $\Vref (R)$ curve based on $\dplx$ and $\rgeo$ respectively, taking $\Usun=10\,\kms$ as an example.

The reference longitude velocity $\vlref$ is then derived using Equation~\ref{eq_vlref}, and the longitudinal-velocity residual of each star is $\dvl \equiv v_l - \vlref$. The footprint is divided into $\Delta x\times \Delta y=0.4\,\kpc\times0.4\,\kpc$ pixels. By assigning each pixel the median $\dvl$ of its stars, we obtain the residual field.

\subsection{Power spectral analysis}\label{sect3.2} 

We compute the two-dimensional (2-D) PS of the residual field using a discrete 2-D Fourier transform \citep{Press07}. Set $M$ and $N$ to be the number of grid points in the $\dvl$ map along $X_h$- and $Y_h$-axes, respectively. To reduce spectral leakage and improve frequency resolution, zero-padding is applied by extending the map dimensions to $M'=2M$ and $N'=2N$. The Fourier transform is calculated via
\begin{align}
\overline{\dvl}\left(f_{x}, f_{y}\right)&=\sum_{m=0}^{M'-1}\sum_{n=0}^{N'-1} \dvl'\left(X_{h,m}, Y_{h,n}\right) \exp \left[-2 \pi i\left(f_{x} X_{h,m}+f_{y} Y_{h,n}\right)\right],   \\ \nonumber
\dvl'&\equiv\dvl-\left \langle\dvl\right \rangle ,
\end{align}\label{eq_FT}
where the discrete frequencies $f_x$ and $f_y$ are given by $f_x=k_x/(M'\Delta x)$ with $k_x=-M'/2,...\,,\,M'/2-1$ and $f_y=k_y/(N'\Delta y)$ with $k_y=-N'/2,...\,,\,N'/2-1$, respectively. The $\left \langle\dvl\right \rangle$ is the mean $\dvl$ of the original map \footnote{The operation removes any spatially constant offset, which would otherwise appear as a zero-frequency component and obscure comparison of the non-zero spatial modes; it preserves all spatially varying kinematic features.}. The 2-D PS is then obtained via
\begin{equation}
    P_{\rm{2D}}(f_x,\,f_y) = \frac{\left|\overline{\dvl} (f_x,\,f_y)\right|^2}{M'N'}.
\end{equation}\label{eq_PS2D}

Figure ~\ref{fig_PS2D} illustrates the procedure for constructing and analyzing the residual field, taking $\Usun=10\,\kms$ as an example: from left to right, it displays the observed field $v_l (X_h,Y_h)$, the reference field $\vlref (X_h,Y_h)$ derived from the baseline obtained in Section ~\ref{sect3.1}, and the residual field $\dvl (X_h,Y_h)$, associated with the 2-D PS of $\dvl (X_h,Y_h)$.

To achieve a direct spectral interpretation, we use the magnitude of the spatial-frequency vector, $f=\sqrt{f_x^2 + f_y^2}$, to obtain the one-dimensional (1-D) counterpart of the 2-D PS. The range of this 1-D frequency $f$ is divided into equally spaced intervals, where the bin width is chosen as $\Delta f=(\Delta f_x + \Delta f_y)/2$, with $\Delta f_x$ and $\Delta f_y$ denoting the frequency spacings of the 2-D PS. For the $j$-th frequency bin $\mathcal{B}_j$, the azimuthally averaged 1-D PS is calculated by
\begin{equation}
    P\left(f_j\right)=\frac{1}{\mathcal{N}_{j}} \sum_{\left(f_{x}, f_{y}\right) \in \mathcal{B}_{j}} P_{2 \mathrm{D}}\left(f_{x}, f_{y}\right),
\end{equation}
where $\mathcal{N}_j$ is the number of 2-D Fourier points that fall within the bin. We adopt $\sqrt{P(f)}$, namely the Fourier amplitude as the function of 1-D frequency, for the following analysis. The random uncertainties of $\sqrt{P(f)}$, which are propagated from the random errors of observables, are estimated based on the statistical Monte Carlo simulations under the assumption that the observational uncertainties are distributed normally \footnote{We generate 400 mock realizations to propagate the quoted observational uncertainties, which is achieved by perturbing the observables according to the Gaussian distribution defined by the quoted measurement uncertainties and correlation coefficients. For each realization, the $\sqrt{\rm{P}(f)}$ is recomputed from the perturbed observables.The median and 16th-84th percentile interval of $\sqrt{\rm{P}(f)}$ are used as as the fiducial result and its propagated random uncertainty, respectively.} .

\section{Results}\label{sect4}

\begin{figure}
    \centering
    \includegraphics[width=9cm]{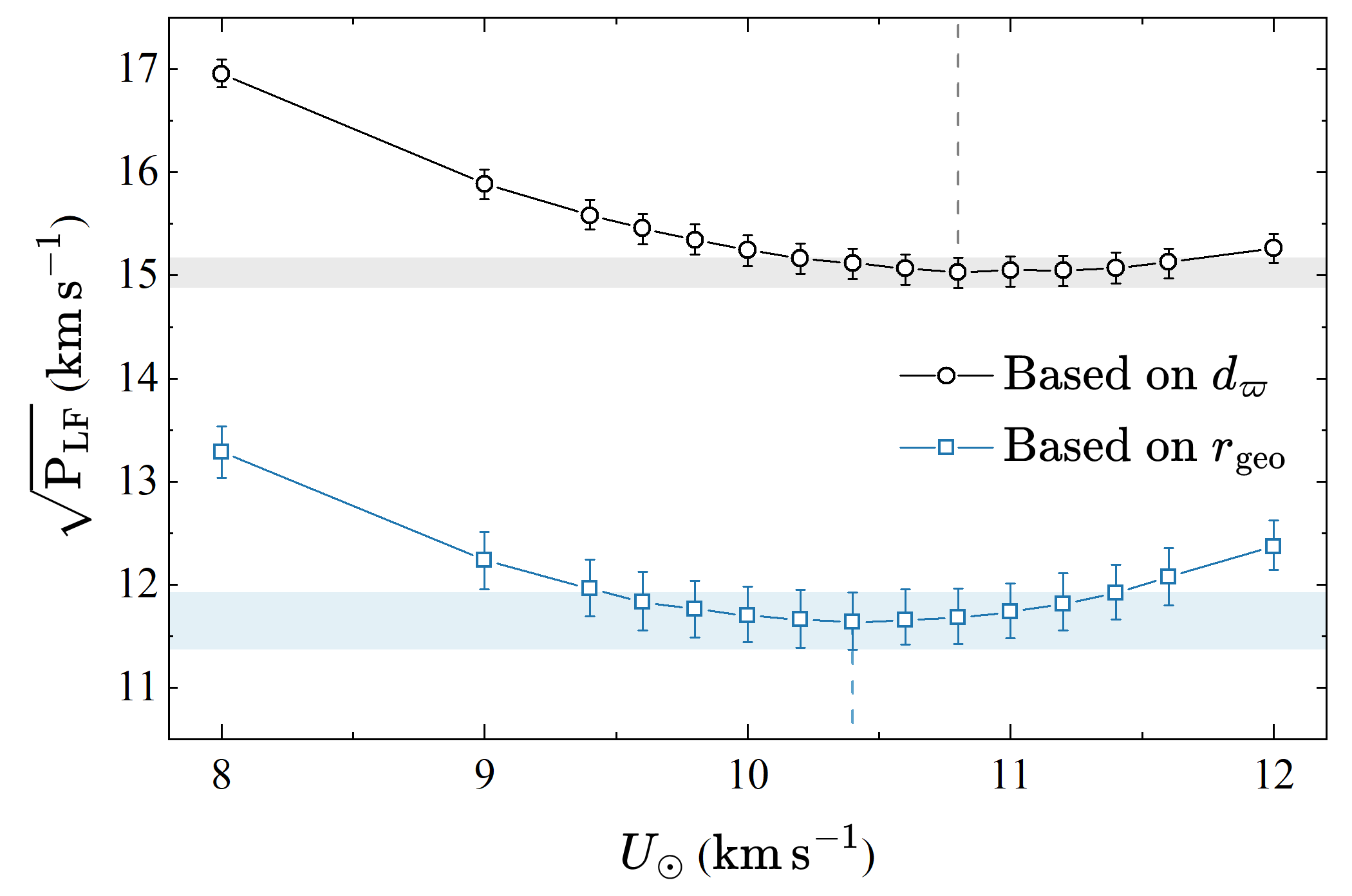}
    \caption{Square root of the mean power at $f<0.2\,\ikpc$ as a function of $\Usun$, for $\dplx$ and $\rgeo$ respectively. A fine sampling step of 0.2 $\kms$ is used within $9.4<\Usun<11.6\,\kms$. The vertical dashed line marks the $\Usun$ value at which the $\sqrtPlf$ reaches its minimum. The horizontal gray band represents the uncertainty interval associated with the minimum value (i.e., the range spanned by the error bars).}
    \label{fig_Plf_Usun}
\end{figure}

Figure \ref{fig_PS1D} presents the amplitude spectra of the $\dvl$ field based on $\dplx$ and $\rgeo$ respectively, assuming different $\Usun$ values for the reference baseline. The 99th-percentile amplitude of the noise spectrum, which is contributed by the random observational uncertainties \footnote{A noise spectrum is obtained from a noise field, which is defined as the difference between the mock $v_l$ and the observed $v_l$.}, is also presented for comparison. The residual amplitudes consistently exceed the noise threshold, indicating that the residuals are statistically significant.

At $f\gtrsim0.54\,\ikpc$, the amplitude slowly increases toward lower frequencies (i.e., larger spatial scales) and remains below $3\,\kms$. The spectra obtained with different $\Usun$ exhibit consistency in this range.

The increase in $\sqrtP$ with decreasing $f$ (i.e., increasing spatial scale) becomes pronounced when $f<0.54\,\ikpc$. At $f\simeq0.33\,\ikpc$, corresponding to a spatial scale of $1/f\simeq3\,\kpc$, $\sqrtP$ reaches around 5-6 $\kms$ and increases monotonically with the adopted $\Usun$, though with variations much smaller than the overall spectral fluctuation. Such a behavior indicates that the residuals have a notable component associated with Galactocentric radial-motion projection. Local stellar streaming motion subject to periodic perturbation from spiral arms (\citealt{Shu16}; \citealt{Bobylev2021}), which is most pronounced along the radial direction, is a plausible contributor. High-dimensional phase-space analysis will enable a more rigorous assessment of this association.

A broad peak is found at $f\simeq0.12$-$0.22\,\ikpc$, corresponding to a scale of approximately 5-8 kpc; the associated amplitude reaches $\gtrsim12\,\kms$. 

The choice of $\Usun$ contributes coherently to the residual power at low frequency, as a biased $\Usun$ produces a dipolar component in the residual field. Consequently, the low-frequency modes can be used to constrain $\Usun$ in a global sense: the optimal value of $\Usun$ minimizes the low-frequency residual power. To quantify this, we define the low-frequency statistic $P_\mathrm{LF}(\Usun)$ to be the mean power over $f<0.2\,\ikpc$, evaluated as a function of $\Usun$. Figure ~\ref{fig_Plf_Usun} shows the estimated $\sqrtPlf$ for each selected $\Usun$; specifically, within the interval where the minimum is attained, we compute the reference field and the corresponding residual $\sqrtPlf$ at $\Usun$ increments of $0.2\,\kms$. 

The $\sqrtPlf$ reaches its minimum at $\Usun=10.8\,\kms$ for $\dplx$ and at $\Usun=10.4\,\kms$ for $\rgeo$. Using $\vsun$ and the derived $\Usun$, the total solar planar velocity with respect to the GC, $S^{(\parallel)}_{\rm{GC}}$, and its apex $\ell_{S,\rm{GC}}$ are as follows:
\begin{equation}\label{eq_solarV1}
    S^{(\parallel)}_{\rm{GC}}= 251.2\,\kms,\;\ell_{S,\rm{GC}}=87.5^\circ
\end{equation}
for $\dplx$, and
\begin{equation}\label{eq_solarV2}
    S^{(\parallel)}_{\rm{GC}}= 251.2\,\kms,\;\ell_{S,\rm{GC}}=87.6^\circ
\end{equation}
for $\rgeo$.

\section{Discussion}
\label{sect5}

\begin{figure}
    \centering
    \includegraphics[width=0.9\linewidth]{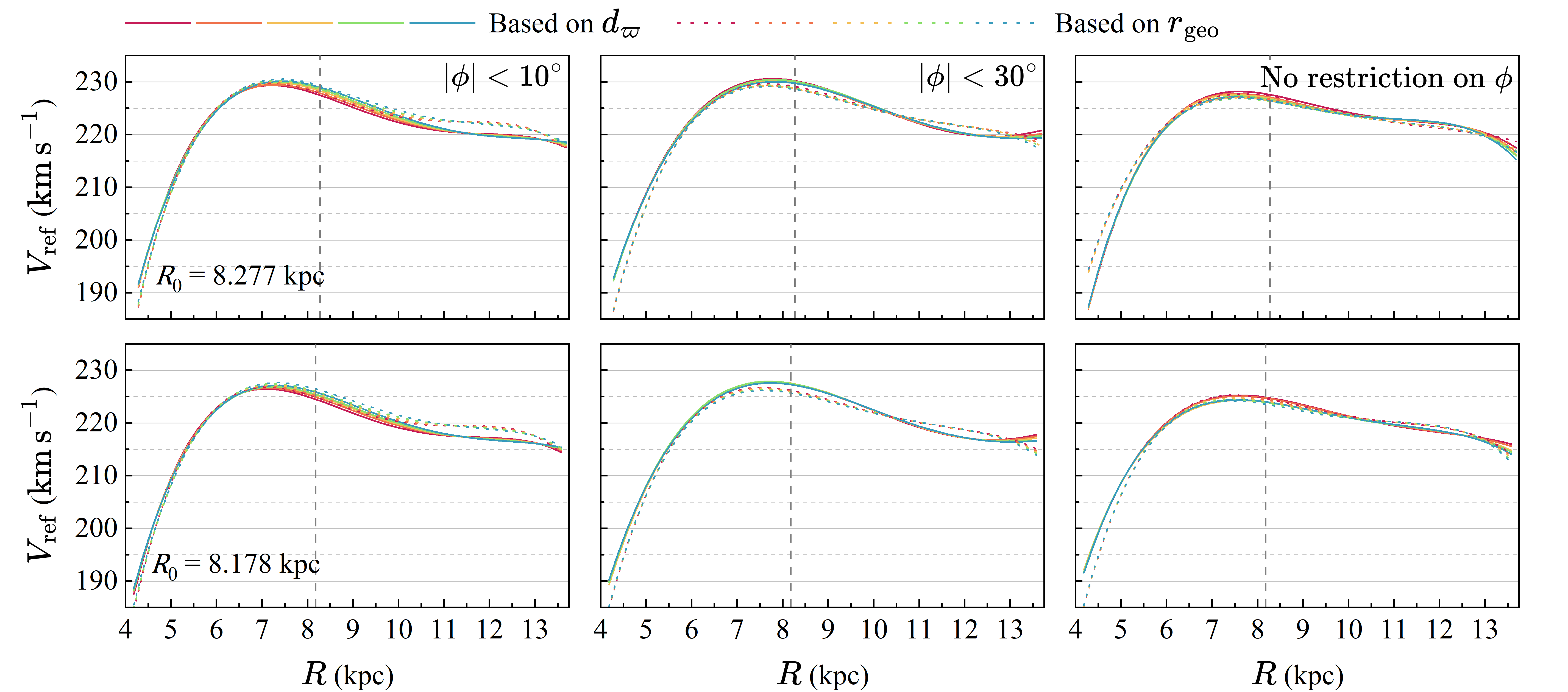}
    \caption{Reference baseline $\Vref (R)$ constructed by varying $R_0$, restriction on $\phi$, and $\Usun$, with $\Usun$ color-coded as in Fig. ~\ref{fig_PS1D}. The top panels display the $\Vref (R)$ obtained assuming $R_0=8.277\,\kpc$, whereas the bottom panels correspond to $R_0=8.178\,\kpc$ (marked by the vertical dashed lines). The left and the middle panels display the $\Vref (R)$ derived from stars within $|\phi|<10^\circ$ and $|\phi|<30^\circ$ respectively; the right panels show the $\Vref (R)$ obtained with no $\phi$ restriction.}
    \label{fig_Vref_byR_byphi}
\end{figure}

\begin{figure}
    \centering
    \includegraphics[width=0.9\linewidth]{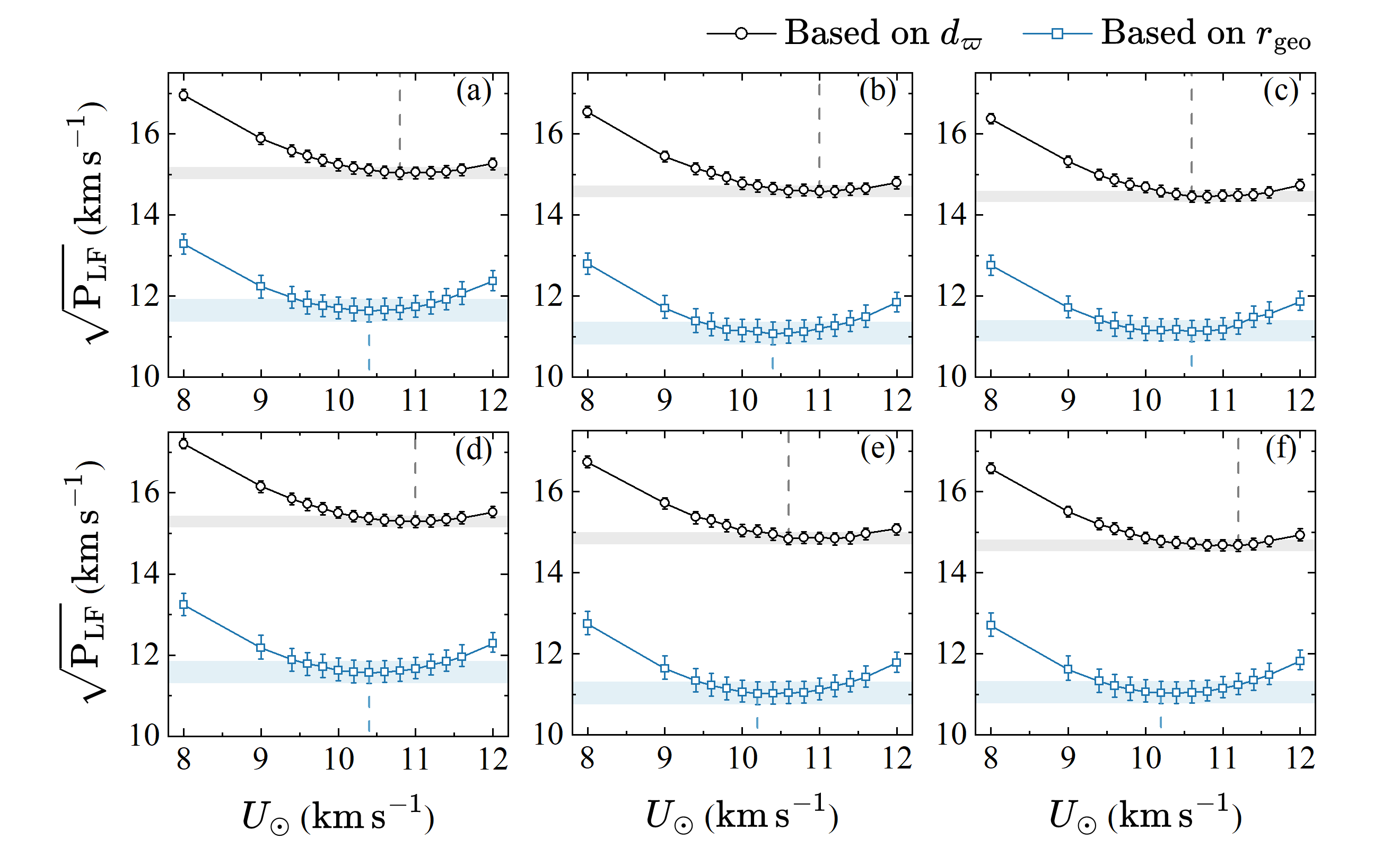}
    \caption{As Fig. ~\ref{fig_Plf_Usun}, but using different reference baselines. Top : $R_0=8.277\,\kpc$; bottom: $R_0=8.178\,\kpc$. Left/middle: baselines from stars with $|\phi| < 10^\circ$ and $|\phi| < 30^\circ$, respectively; right: baselines with no $\phi$ restriction.}
    \label{fig_Plf_Usun_byR_byphi}
\end{figure}

\begin{figure}
    \centering
    \includegraphics[width=1\linewidth]{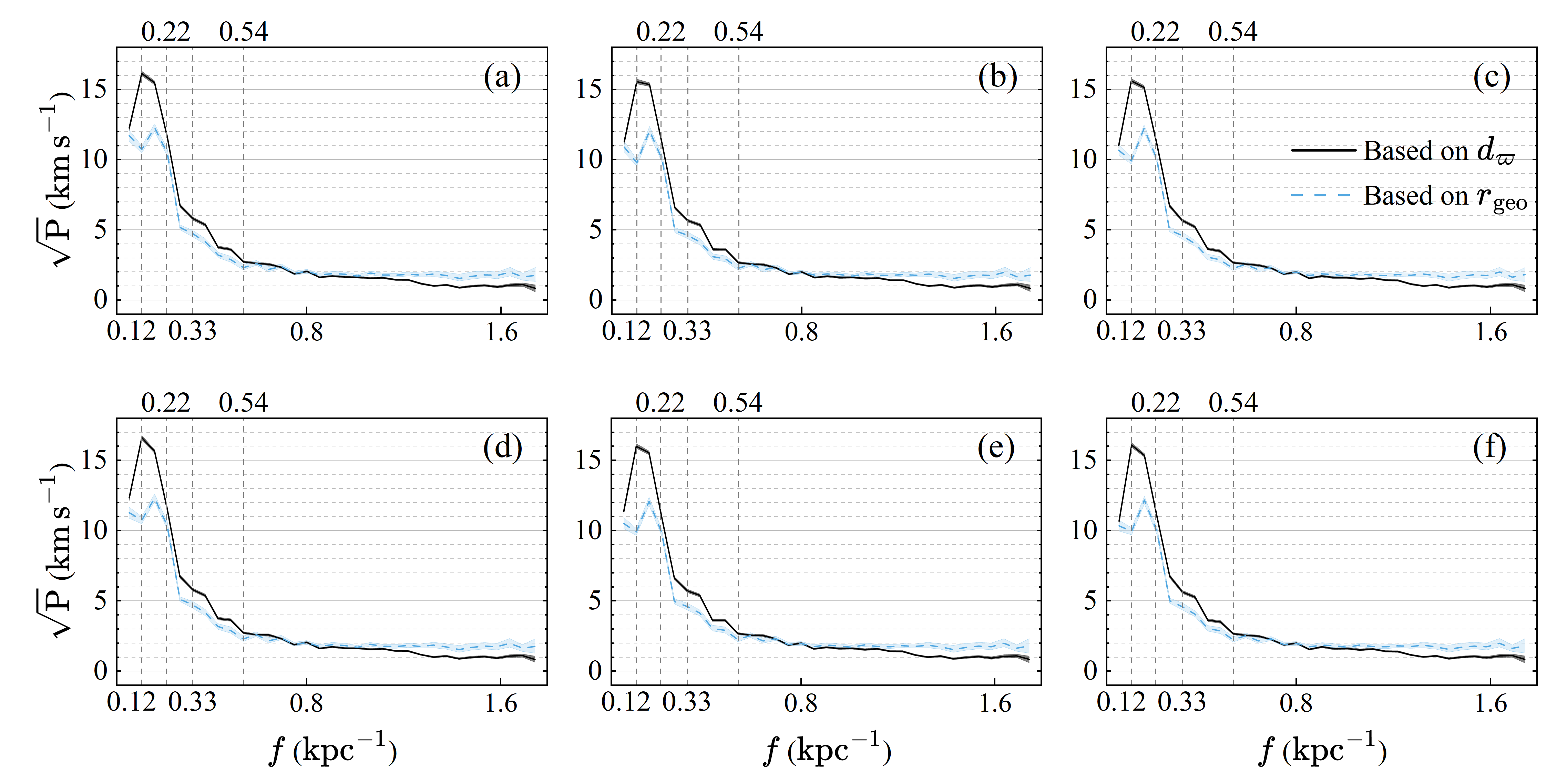}
    \caption{As panels (a1) and (b1) of Fig. ~\ref{fig_PS1D}, but based on distinct reference baselines: for each spectrum, the $R_0$ and the $\phi$ range adopted in constructing $\Vref (R)$ are identical to those in Fig. ~\ref{fig_Plf_Usun_byR_byphi}, and the $\Usun$ is adopted as the preferred value found in Fig. ~\ref{fig_Plf_Usun_byR_byphi}. Vertical dashed lines mark several characteristic frequencies, with labels alternating between the top and bottom axes of each panel for clarity. }
    \label{fig_PS1D_byR_byphi}
\end{figure}

Using the power spectral analysis, we identify the characteristic spatial scales corresponding to non-axisymmetric modes in the residual velocity field. In addition, we derive a global estimate of $\Usun$ that is decoupled from local streaming motions. Equations \eqref{eq_solarV1} and \eqref{eq_solarV2} thus provide independent kinematic determinations of the solar planar velocity in the Galactocentric rest frame. Meanwhile, due to the observational random uncertainties propagated into the analysis, the $\Usun$ values of $10.2$-$11.6\,\kms$ for $\dplx$ and of $9.6$-$11.4\,\kms$ for $\rgeo$ remain statistically consistent with each other within their respective error range (see Fig. ~\ref{fig_Plf_Usun}). The precision of the estimation is expected to improve in future analysis with expanded dataset, and its robustness will be further assessed using independent samples with distinct evolutionary stages or spatial footprints. 

Focusing on the current sample, variations in the adopted $R_0$ (specifically, substituting $R_0=8.178\,\kpc$ \citep{Gravity19}) or modifications to the $\phi$ restriction for constructing the baseline (either using $|\phi|<30^\circ$ or removing the $\phi$ cut) induce changes in $\Vref (R)$ of a few $\kms$. These perturbations lead to only minor adjustments in the derived spectrum slightly and preserve the preferred $\Usun$ within uncertainties. Figure ~\ref{fig_Vref_byR_byphi} presents the baseline curves derived under alternate conditions; Figure ~\ref{fig_Plf_Usun_byR_byphi} shows the corresponding $\sqrtPlf$ with $\Usun$.

Figure ~\ref{fig_PS1D_byR_byphi} presents the spectra computed using the alternate $\Vref (R)$; the $\Usun$ used for constructing each $\Vref (R)$ is the preferred value found in Fig. ~\ref{fig_Plf_Usun_byR_byphi}. The resulting amplitude profiles of the residual field are generally consistent across different baselines. Having minimized the large-scale residual signal produced by a biased $\Usun$, the remaining residual amplitudes shown by Fig. ~\ref{fig_PS1D_byR_byphi} suggest genuine non-axisymmetric structures in the disk kinematics under the assumed framework. The most pronounced signal with scales of $\sim5$-$8\,\kpc$ suggests a disk-wide mode that may contain responses to the bar, the projection of a triaxial halo (\citealt{KT94}; \citealt{Bovy15}), or other specific perturber. With amplitudes $>12\,\kms$, this signal provides a quantitative diagnostic that can be compared with dynamical models and simulations. 

Meanwhile, the measured residual field may additionally contain unmodeled systematics in the data, including spatially correlated astrometric errors \citep{Lindegren21b} and conditional distance scales. The discrepancy between the amplitudes based on $\dplx$ and $\rgeo$ indicates that such systematics may introduce a velocity amplitude bias of a few $\kms$ for structures on scales $>5\,\kpc$. Moreover, the finite spatial window can induce spectral leakage among the lowest-frequency mode. The upcoming Gaia DR4 will improve astrometric precision and extend distance coverage, thus improving sensitivity to disk-wide Galactic structures. The data systematics -- especially the distance scale calibration -- need to be characterized rigorously before drawing specific dynamical interpretations. 

\section{Conclusions}\label{sect6}

Employing nearly four million Gaia DR3 RGB stars with high-quality astrometry, we have measured the spatial PS of Galactic longitudinal-velocity residuals over an $8\,\kpc\times12\,\kpc$ heliocentric region. The power is significant from random noise based both on $\dplx$ and $\rgeo$. The spectra exhibit a pronounced broad peak at $f\simeq0.12$-$0.22\,\ikpc$ (corresponding to a scale $\simeq5$-$8\,\kpc$). Taking into account the variations of the solar radial velocity used in constructing the reference baseline, there is a notable spectral signal associated with a candidate radial streaming motion at $f\simeq0.33\,\ikpc$ (corresponding to a scale $\simeq3\,\kpc$). The low-frequency power minimum yields a global kinematic constraint on the solar radial motion, namely $\Usun\simeq10.8\,\kms$ and $10.4\,\kms$ for the two distance scales respectively. Under the assumption of $R_0=8.277\,\kpc$, this constraint thus yields an estimate of the Sun's total planar velocity relative to the GC: $251.2\,\kms$, with the apex (i.e., the Galactic longitude of the velocity vector) of $87.5^\circ$-$87.6^\circ$. Based on the preferred $\Usun$, the maximum amplitude of the residuals reaches approximately $16\,\kms$ for $\dplx$ and $12\,\kms$ for $\rgeo$. The potential-free PS provides an empirical target for dynamical modeling and simulations.

\begin{acknowledgements}
This work has made use of data from the European Space Agency (ESA) mission {\it Gaia} (\url{https://www.cosmos.esa.int/gaia}), processed by the {\it Gaia} Data Processing and Analysis Consortium (DPAC, \url{https://www.cosmos.esa.int/web/gaia/dpac/consortium}). Funding for the DPAC has been provided by national institutions, in particular the institutions participating in the {\it Gaia} Multilateral Agreement. This work is supported by the National Key Research and Development Program of China No. 2022YFA1603204. The author thanks Prof. Jia-Cheng Liu for helpful advice on the presentation and clarity of this manuscript. 
\end{acknowledgements}


\bibliographystyle{raa}
\bibliography{references}

@article{Lindegren21b,
  author = {{Lindegren}, L. and {Bastian}, U. and {Biermann}, M. and {Bombrun}, A. and {de Torres}, A. and {Gerlach}, E. and {Geyer}, R. and {Hern{\'a}ndez}, J. and {Hilger}, T. and {Hobbs}, D. and {Klioner}, S.~A. and {Lammers}, U. and {McMillan}, P.~J. and {Ramos-Lerate}, M. and {Steidelm{\"u}ller}, H. and {Stephenson}, C.~A. and {van Leeuwen}, F.},
  title = {{Gaia Early Data Release 3. Parallax bias versus magnitude, colour, and position}},
  journal = {\aap},
  year = {2021b},
  month = may,
  volume = {649},
  eid = {A4},
  pages = {A4},
  doi = {10.1051/0004-6361/202039653},
  archiveprefix = {arXiv},
  eprint = {2012.01742},
  primaryclass = {astro-ph.IM},
  adsurl = {https://ui.adsabs.harvard.edu/abs/2021A&A...649A...4L},
}

@article{Lindegren21a,
  author = {{Lindegren}, L. and {Klioner}, S.~A. and {Hern{\'a}ndez}, J. and {Bombrun}, A. and {Ramos-Lerate}, M. and {Steidelm{\"u}ller}, H. and {Bastian}, U. and {Biermann}, M. and {de Torres}, A. and {Gerlach}, E. and {Geyer}, R. and {Hilger}, T. and {Hobbs}, D. and {Lammers}, U. and {McMillan}, P.~J. and {Stephenson}, C.~A. and {Casta{\~n}eda}, J. and {Davidson}, M. and {Fabricius}, C. and {Gracia-Abril}, G. and {Portell}, J. and {Rowell}, N. and {Teyssier}, D. and {Torra}, F. and {Bartolom{\'e}}, S. and {Clotet}, M. and {Garralda}, N. and {Gonz{\'a}lez-Vidal}, J.~J. and {Torra}, J. and {Abbas}, U. and {Altmann}, M. and {Anglada Varela}, E. and {Balaguer-N{\'u}{\~n}ez}, L. and {Balog}, Z. and {Barache}, C. and {Becciani}, U. and {Bernet}, M. and {Bertone}, S. and {Bianchi}, L. and {Bouquillon}, S. and {Brown}, A.~G.~A. and {Bucciarelli}, B. and {Busonero}, D. and {Butkevich}, A.~G. and {Buzzi}, R. and {Cancelliere}, R. and {Carlucci}, T. and {Charlot}, P. and {Cioni}, M.-R.~L. and {Crosta}, M. and {Crowley}, C. and {del Peloso}, E.~F. and {del Pozo}, E. and {Drimmel}, R. and {Esquej}, P. and {Fienga}, A. and {Fraile}, E. and {Gai}, M. and {Garcia-Reinaldos}, M. and {Guerra}, R. and {Hambly}, N.~C. and {Hauser}, M. and {Jan{\ss}en}, K. and {Jordan}, S. and {Kostrzewa-Rutkowska}, Z. and {Lattanzi}, M.~G. and {Liao}, S. and {Licata}, E. and {Lister}, T.~A. and {L{\"o}ffler}, W. and {Marchant}, J.~M. and {Masip}, A. and {Mignard}, F. and {Mints}, A. and {Molina}, D. and {Mora}, A. and {Morbidelli}, R. and {Murphy}, C.~P. and {Pagani}, C. and {Panuzzo}, P. and {Pe{\~n}alosa Esteller}, X. and {Poggio}, E. and {Re Fiorentin}, P. and {Riva}, A. and {Sagrist{\`a} Sell{\'e}s}, A. and {Sanchez Gimenez}, V. and {Sarasso}, M. and {Sciacca}, E. and {Siddiqui}, H.~I. and {Smart}, R.~L. and {Souami}, D. and {Spagna}, A. and {Steele}, I.~A. and {Taris}, F. and {Utrilla}, E. and {van Reeven}, W. and {Vecchiato}, A.},
  title = {{Gaia Early Data Release 3. The astrometric solution}},
  journal = {\aap},
  year = {2021a},
  month = may,
  volume = {649},
  eid = {A2},
  pages = {A2},
  doi = {10.1051/0004-6361/202039709},
  archiveprefix = {arXiv},
  eprint = {2012.03380},
  primaryclass = {astro-ph.IM},
  adsurl = {https://ui.adsabs.harvard.edu/abs/2021A&A...649A...2L},
}

@article{Eilers20,
  author = {{Eilers}, Anna-Christina and {Hogg}, David W. and {Rix}, Hans-Walter and {Frankel}, Neige and {Hunt}, Jason A.~S. and {Fouvry}, Jean-Baptiste and {Buck}, Tobias},
  title = {{The Strength of the Dynamical Spiral Perturbation in the Galactic Disk}},
  journal = {\apj},
  year = {2020},
  month = sep,
  volume = {900},
  number = {2},
  eid = {186},
  pages = {186},
  doi = {10.3847/1538-4357/abac0b},
  archiveprefix = {arXiv},
  eprint = {2003.01132},
  primaryclass = {astro-ph.GA},
  adsurl = {https://ui.adsabs.harvard.edu/abs/2020ApJ...900..186E},
}

@article{Lin26,
  author = {{Lin}, Z.~H. and {Hao}, C.~J. and {Liu}, D.~J. and {Li}, Y.~J. and {Li}, J.~J. and {Dong}, Y.~W. and {He}, Y.~X. and {Esimbek}, Jarken},
  title = {{Large-scale Spiral Structure of the Milky Way Traced by Young Classical Cepheids}},
  journal = {\apjl},
  year = {2026},
  month = oct,
  volume = {1009},
  number = {2},
  eid = {L33},
  pages = {L33},
  doi = {10.3847/2041-8213/aea61c},
  archiveprefix = {arXiv},
  eprint = {2609.34208},
  primaryclass = {astro-ph.GA},
  adsurl = {https://ui.adsabs.harvard.edu/abs/2026ApJ..1009L..33L},
}

@article{YB26,
  author = {{Yamaguchi}, Tomoki and {Baba}, Junichi},
  title = {{Axisymmetric Jeans modelling systematically overestimates the circular speed in the inner Milky Way}},
  journal = {\pasj},
  year = {2026},
  month = aug,
  volume = {78},
  number = {4},
  pages = {1363-1371},
  doi = {10.1093/pasj/psag066},
  archiveprefix = {arXiv},
  eprint = {2605.09373},
  primaryclass = {astro-ph.GA},
  adsurl = {https://ui.adsabs.harvard.edu/abs/2026PASJ...78.1363Y},
}

@article{Fedorov26,
  author = {{Fedorov}, P.~N. and {Dmytrenko}, A.~M. and {Akhmetov}, V.~S. and {Velichko}, A.~B. and {Zheleznyak}, A.~P. and {Khramtsov}, V.~P. and {Denyshchenko}, S.~I. and {Vavilova}, I.~B. and {Dobrycheva}, D.~V. and {Sergijenko}, O. and {Vasylenko}, A.~A. and {Kompaniiets}, O.~V.},
  title = {{Local Jeans estimates of the Galaxy's circular velocity, based on RGB stars from the Gaia catalogue}},
  journal = {\mnras},
  year = {2026},
  month = aug,
  volume = {550},
  number = {4},
  eid = {stag1310},
  pages = {stag1310},
  doi = {10.1093/mnras/stag1310},
  archiveprefix = {arXiv},
  eprint = {2511.22295},
  primaryclass = {astro-ph.GA},
  adsurl = {https://ui.adsabs.harvard.edu/abs/2026MNRAS.550g1310F},
}

@article{BB23,
  author = {Bobylev, V. V. and Bajkova, A. T.},
  title = {3D Kinematics of Classical Cepheids According to Gaia EDR3 Catalog},
  journal = {Research in Astronomy and Astrophysics},
  year = {2023},
  volume = {23},
  issn = {2397-6209},
  url = {http://dx.doi.org/10.1088/1674-4527/acb878},
  doi = {10.1088/1674-4527/acb878},
  number = {4},
  publisher = {IOP Publishing},
  month = Mar,
  pages = {045001},
}

@article{Bailer-Jones21,
  author = {{Bailer-Jones}, C.~A.~L. and {Rybizki}, J. and {Fouesneau}, M. and {Demleitner}, M. and {Andrae}, R.},
  title = {{Estimating Distances from Parallaxes. V. Geometric and Photogeometric Distances to 1.47 Billion Stars in Gaia Early Data Release 3}},
  journal = {\aj},
  year = {2021},
  month = mar,
  volume = {161},
  number = {3},
  eid = {147},
  pages = {147},
  doi = {10.3847/1538-3881/abd806},
  archiveprefix = {arXiv},
  eprint = {2012.05220},
  primaryclass = {astro-ph.SR},
  adsurl = {https://ui.adsabs.harvard.edu/abs/2021AJ....161..147B},
}

@article{Gaia23b,
  author = {{{Gaia Collaboration}} and {Drimmel}, R. and {Romero-G{\'o}mez}, M. and {Chemin}, L. and {Ramos}, P. and {Poggio}, E. and {Ripepi}, V. and {Andrae}, R. and {Blomme}, R. and {Cantat-Gaudin}, T. and {Castro-Ginard}, A. and {Clementini}, G. and {Figueras}, F. and {Fouesneau}, M. and {Fr{\'e}mat}, Y. and {Jardine}, K. and {Khanna}, S. and {Lobel}, A. and {Marshall}, D.~J. and {Muraveva}, T. and {Brown}, A.~G.~A. and {Vallenari}, A. and {Prusti}, T. and {de Bruijne}, J.~H.~J. and {Arenou}, F. and {Babusiaux}, C. and {Biermann}, M. and {Creevey}, O.~L. and {Ducourant}, C. and {Evans}, D.~W. and {Eyer}, L. and {Guerra}, R. and {Hutton}, A. and {Jordi}, C. and {Klioner}, S.~A. and {Lammers}, U.~L. and {Lindegren}, L. and {Luri}, X. and {Mignard}, F. and {Panem}, C. and {Pourbaix}, D. and {Randich}, S. and {Sartoretti}, P. and {Soubiran}, C. and {Tanga}, P. and {Walton}, N.~A. and {Bailer-Jones}, C.~A.~L. and {Bastian}, U. and {Jansen}, F. and {Katz}, D. and {Lattanzi}, M.~G. and {van Leeuwen}, F. and {Bakker}, J. and {Cacciari}, C. and {Casta{\~n}eda}, J. and {De Angeli}, F. and {Fabricius}, C. and {Galluccio}, L. and {Guerrier}, A. and {Heiter}, U. and {Masana}, E. and {Messineo}, R. and {Mowlavi}, N. and {Nicolas}, C. and {Nienartowicz}, K. and {Pailler}, F. and {Panuzzo}, P. and {Riclet}, F. and {Roux}, W. and {Seabroke}, G.~M. and {Sordo}, R. and {Th{\'e}venin}, F. and {Gracia-Abril}, G. and {Portell}, J. and {Teyssier}, D. and {Altmann}, M. and {Audard}, M. and {Bellas-Velidis}, I. and {Benson}, K. and {Berthier}, J. and {Burgess}, P.~W. and {Busonero}, D. and {Busso}, G. and {C{\'a}novas}, H. and {Carry}, B. and {Cellino}, A. and {Cheek}, N. and {Damerdji}, Y. and {Davidson}, M. and {de Teodoro}, P. and {Nu{\~n}ez Campos}, M. and {Delchambre}, L. and {Dell'Oro}, A. and {Esquej}, P. and {Fern{\'a}ndez-Hern{\'a}ndez}, J. and {Fraile}, E. and {Garabato}, D. and {Garc{\'\i}a-Lario}, P. and {Gosset}, E. and {Haigron}, R. and {Halbwachs}, J.-L. and {Hambly}, N.~C. and {Harrison}, D.~L. and {Hern{\'a}ndez}, J. and {Hestroffer}, D. and {Hodgkin}, S.~T. and {Holl}, B. and {Jan{\ss}en}, K. and {Jevardat de Fombelle}, G. and {Jordan}, S. and {Krone-Martins}, A. and {Lanzafame}, A.~C. and {L{\"o}ffler}, W. and {Marchal}, O. and {Marrese}, P.~M. and {Moitinho}, A. and {Muinonen}, K. and {Osborne}, P. and {Pancino}, E. and {Pauwels}, T. and {Recio-Blanco}, A. and {Reyl{\'e}}, C. and {Riello}, M. and {Rimoldini}, L. and {Roegiers}, T. and {Rybizki}, J. and {Sarro}, L.~M. and {Siopis}, C. and {Smith}, M. and {Sozzetti}, A. and {Utrilla}, E. and {van Leeuwen}, M. and {Abbas}, U. and {{\'A}brah{\'a}m}, P. and {Abreu Aramburu}, A. and {Aerts}, C. and {Aguado}, J.~J. and {Ajaj}, M. and {Aldea-Montero}, F. and {Altavilla}, G. and {{\'A}lvarez}, M.~A. and {Alves}, J. and {Anders}, F. and {Anderson}, R.~I. and {Anglada Varela}, E. and {Antoja}, T. and {Baines}, D. and {Baker}, S.~G. and {Balaguer-N{\'u}{\~n}ez}, L. and {Balbinot}, E. and {Balog}, Z. and {Barache}, C. and {Barbato}, D. and {Barros}, M. and {Barstow}, M.~A. and {Bartolom{\'e}}, S. and {Bassilana}, J.-L. and {Bauchet}, N. and {Becciani}, U. and {Bellazzini}, M. and {Berihuete}, A. and {Bernet}, M. and {Bertone}, S. and {Bianchi}, L. and {Binnenfeld}, A. and {Blanco-Cuaresma}, S. and {Boch}, T. and {Bombrun}, A. and {Bossini}, D. and {Bouquillon}, S. and {Bragaglia}, A. and {Bramante}, L. and {Breedt}, E. and {Bressan}, A. and {Brouillet}, N. and {Brugaletta}, E. and {Bucciarelli}, B. and {Burlacu}, A. and {Butkevich}, A.~G. and {Buzzi}, R. and {Caffau}, E. and {Cancelliere}, R. and {Carballo}, R. and {Carlucci}, T. and {Carnerero}, M.~I. and {Carrasco}, J.~M. and {Casamiquela}, L. and {Castellani}, M. and {Chaoul}, L. and {Charlot}, P. and {Chiaramida}, V. and {Chiavassa}, A. and {Chornay}, N. and {Comoretto}, G. and {Contursi}, G. and {Cooper}, W.~J. and {Cornez}, T. and {Cowell}, S. and {Crifo}, F. and {Cropper}, M.},
  title = {{Gaia Data Release 3. Mapping the asymmetric disc of the Milky Way}},
  journal = {\aap},
  year = {2023b},
  month = jun,
  volume = {674},
  eid = {A37},
  pages = {A37},
  doi = {10.1051/0004-6361/202243797},
  archiveprefix = {arXiv},
  eprint = {2206.06207},
  primaryclass = {astro-ph.GA},
  adsurl = {https://ui.adsabs.harvard.edu/abs/2023A&A...674A..37G},
}

@article{Gaia23a,
  author = {{Gaia Collaboration} and {Vallenari}, A. and {Brown}, A.~G.~A. and {Prusti}, T. and {de Bruijne}, J.~H.~J. and {Arenou}, F. and {Babusiaux}, C. and {Biermann}, M. and {Creevey}, O.~L. and {Ducourant}, C. and {Evans}, D.~W. and {Eyer}, L. and {Guerra}, R. and {Hutton}, A. and {Jordi}, C. and {Klioner}, S.~A. and {Lammers}, U.~L. and {Lindegren}, L. and {Luri}, X. and {Mignard}, F. and {Panem}, C. and {Pourbaix}, D. and {Randich}, S. and {Sartoretti}, P. and {Soubiran}, C. and {Tanga}, P. and {Walton}, N.~A. and {Bailer-Jones}, C.~A.~L. and {Bastian}, U. and {Drimmel}, R. and {Jansen}, F. and {Katz}, D. and {Lattanzi}, M.~G. and {van Leeuwen}, F. and {Bakker}, J. and {Cacciari}, C. and {Casta{\~n}eda}, J. and {De Angeli}, F. and {Fabricius}, C. and {Fouesneau}, M. and {Fr{\'e}mat}, Y. and {Galluccio}, L. and {Guerrier}, A. and {Heiter}, U. and {Masana}, E. and {Messineo}, R. and {Mowlavi}, N. and {Nicolas}, C. and {Nienartowicz}, K. and {Pailler}, F. and {Panuzzo}, P. and {Riclet}, F. and {Roux}, W. and {Seabroke}, G.~M. and {Sordo}, R. and {Th{\'e}venin}, F. and {Gracia-Abril}, G. and {Portell}, J. and {Teyssier}, D. and {Altmann}, M. and {Andrae}, R. and {Audard}, M. and {Bellas-Velidis}, I. and {Benson}, K. and {Berthier}, J. and {Blomme}, R. and {Burgess}, P.~W. and {Busonero}, D. and {Busso}, G. and {C{\'a}novas}, H. and {Carry}, B. and {Cellino}, A. and {Cheek}, N. and {Clementini}, G. and {Damerdji}, Y. and {Davidson}, M. and {de Teodoro}, P. and {Nu{\~n}ez Campos}, M. and {Delchambre}, L. and {Dell'Oro}, A. and {Esquej}, P. and {Fern{\'a}ndez-Hern{\'a}ndez}, J. and {Fraile}, E. and {Garabato}, D. and {Garc{\'\i}a-Lario}, P. and {Gosset}, E. and {Haigron}, R. and {Halbwachs}, J.-L. and {Hambly}, N.~C. and {Harrison}, D.~L. and {Hern{\'a}ndez}, J. and {Hestroffer}, D. and {Hodgkin}, S.~T. and {Holl}, B. and {Jan{\ss}en}, K. and {Jevardat de Fombelle}, G. and {Jordan}, S. and {Krone-Martins}, A. and {Lanzafame}, A.~C. and {L{\"o}ffler}, W. and {Marchal}, O. and {Marrese}, P.~M. and {Moitinho}, A. and {Muinonen}, K. and {Osborne}, P. and {Pancino}, E. and {Pauwels}, T. and {Recio-Blanco}, A. and {Reyl{\'e}}, C. and {Riello}, M. and {Rimoldini}, L. and {Roegiers}, T. and {Rybizki}, J. and {Sarro}, L.~M. and {Siopis}, C. and {Smith}, M. and {Sozzetti}, A. and {Utrilla}, E. and {van Leeuwen}, M. and {Abbas}, U. and {{\'A}brah{\'a}m}, P. and {Abreu Aramburu}, A. and {Aerts}, C. and {Aguado}, J.~J. and {Ajaj}, M. and {Aldea-Montero}, F. and {Altavilla}, G. and {{\'A}lvarez}, M.~A. and {Alves}, J. and {Anders}, F. and {Anderson}, R.~I. and {Anglada Varela}, E. and {Antoja}, T. and {Baines}, D. and {Baker}, S.~G. and {Balaguer-N{\'u}{\~n}ez}, L. and {Balbinot}, E. and {Balog}, Z. and {Barache}, C. and {Barbato}, D. and {Barros}, M. and {Barstow}, M.~A. and {Bartolom{\'e}}, S. and {Bassilana}, J.-L. and {Bauchet}, N. and {Becciani}, U. and {Bellazzini}, M. and {Berihuete}, A. and {Bernet}, M. and {Bertone}, S. and {Bianchi}, L. and {Binnenfeld}, A. and {Blanco-Cuaresma}, S. and {Blazere}, A. and {Boch}, T. and {Bombrun}, A. and {Bossini}, D. and {Bouquillon}, S. and {Bragaglia}, A. and {Bramante}, L. and {Breedt}, E. and {Bressan}, A. and {Brouillet}, N. and {Brugaletta}, E. and {Bucciarelli}, B. and {Burlacu}, A. and {Butkevich}, A.~G. and {Buzzi}, R. and {Caffau}, E. and {Cancelliere}, R. and {Cantat-Gaudin}, T. and {Carballo}, R. and {Carlucci}, T. and {Carnerero}, M.~I. and {Carrasco}, J.~M. and {Casamiquela}, L. and {Castellani}, M. and {Castro-Ginard}, A. and {Chaoul}, L. and {Charlot}, P. and {Chemin}, L. and {Chiaramida}, V. and {Chiavassa}, A. and {Chornay}, N. and {Comoretto}, G. and {Contursi}, G. and {Cooper}, W.~J. and {Cornez}, T. and {Cowell}, S. and {Crifo}, F. and {Cropper}, M. and {Crosta}, M. and {Crowley}, C. and {Dafonte}, C. and {Dapergolas}, A. and {David}, M. and {David}, P. and {de Laverny}, P. and {De Luise}, F. and {De March}, R.},
  title = {{Gaia Data Release 3. Summary of the content and survey properties}},
  journal = {\aap},
  year = {2023a},
  month = jun,
  volume = {674},
  eid = {A1},
  pages = {A1},
  doi = {10.1051/0004-6361/202243940},
  archiveprefix = {arXiv},
  eprint = {2208.00211},
  primaryclass = {astro-ph.GA},
  adsurl = {https://ui.adsabs.harvard.edu/abs/2023A&A...674A...1G},
}

@article{Gaia23c,
  author = {{Gaia Collaboration} and {Creevey}, O.~L. and {Sarro}, L.~M. and {Lobel}, A. and {Pancino}, E. and {Andrae}, R. and {Smart}, R.~L. and {Clementini}, G. and {Heiter}, U. and {Korn}, A.~J. and {Fouesneau}, M. and {Fr{\'e}mat}, Y. and {De Angeli}, F. and {Vallenari}, A. and {Harrison}, D.~L. and {Th{\'e}venin}, F. and {Reyl{\'e}}, C. and {Sordo}, R. and {Garofalo}, A. and {Brown}, A.~G.~A. and {Eyer}, L. and {Prusti}, T. and {de Bruijne}, J.~H.~J. and {Arenou}, F. and {Babusiaux}, C. and {Biermann}, M. and {Ducourant}, C. and {Evans}, D.~W. and {Guerra}, R. and {Hutton}, A. and {Jordi}, C. and {Klioner}, S.~A. and {Lammers}, U.~L. and {Lindegren}, L. and {Luri}, X. and {Mignard}, F. and {Panem}, C. and {Pourbaix}, D. and {Randich}, S. and {Sartoretti}, P. and {Soubiran}, C. and {Tanga}, P. and {Walton}, N.~A. and {Bailer-Jones}, C.~A.~L. and {Bastian}, U. and {Drimmel}, R. and {Jansen}, F. and {Katz}, D. and {Lattanzi}, M.~G. and {van Leeuwen}, F. and {Bakker}, J. and {Cacciari}, C. and {Casta{\~n}eda}, J. and {Fabricius}, C. and {Galluccio}, L. and {Guerrier}, A. and {Masana}, E. and {Messineo}, R. and {Mowlavi}, N. and {Nicolas}, C. and {Nienartowicz}, K. and {Pailler}, F. and {Panuzzo}, P. and {Riclet}, F. and {Roux}, W. and {Seabroke}, G.~M. and {Gracia-Abril}, G. and {Portell}, J. and {Teyssier}, D. and {Altmann}, M. and {Audard}, M. and {Bellas-Velidis}, I. and {Benson}, K. and {Berthier}, J. and {Blomme}, R. and {Burgess}, P.~W. and {Busonero}, D. and {Busso}, G. and {C{\'a}novas}, H. and {Carry}, B. and {Cellino}, A. and {Cheek}, N. and {Damerdji}, Y. and {Davidson}, M. and {de Teodoro}, P. and {Nu{\~n}ez Campos}, M. and {Delchambre}, L. and {Dell'Oro}, A. and {Esquej}, P. and {Fern{\'a}ndez-Hern{\'a}ndez}, J. and {Fraile}, E. and {Garabato}, D. and {Garc{\'\i}a-Lario}, P. and {Gosset}, E. and {Haigron}, R. and {Halbwachs}, J.-L. and {Hambly}, N.~C. and {Hern{\'a}ndez}, J. and {Hestroffer}, D. and {Hodgkin}, S.~T. and {Holl}, B. and {Jan{\ss}en}, K. and {Jevardat de Fombelle}, G. and {Jordan}, S. and {Krone-Martins}, A. and {Lanzafame}, A.~C. and {L{\"o}ffler}, W. and {Marchal}, O. and {Marrese}, P.~M. and {Moitinho}, A. and {Muinonen}, K. and {Osborne}, P. and {Pauwels}, T. and {Recio-Blanco}, A. and {Riello}, M. and {Rimoldini}, L. and {Roegiers}, T. and {Rybizki}, J. and {Siopis}, C. and {Smith}, M. and {Sozzetti}, A. and {Utrilla}, E. and {van Leeuwen}, M. and {Abbas}, U. and {{\'A}brah{\'a}m}, P. and {Abreu Aramburu}, A. and {Aerts}, C. and {Aguado}, J.~J. and {Ajaj}, M. and {Aldea-Montero}, F. and {Altavilla}, G. and {{\'A}lvarez}, M.~A. and {Alves}, J. and {Anders}, F. and {Anderson}, R.~I. and {Anglada Varela}, E. and {Antoja}, T. and {Baines}, D. and {Baker}, S.~G. and {Balaguer-N{\'u}{\~n}ez}, L. and {Balbinot}, E. and {Balog}, Z. and {Barache}, C. and {Barbato}, D. and {Barros}, M. and {Barstow}, M.~A. and {Bartolom{\'e}}, S. and {Bassilana}, J.-L. and {Bauchet}, N. and {Becciani}, U. and {Bellazzini}, M. and {Berihuete}, A. and {Bernet}, M. and {Bertone}, S. and {Bianchi}, L. and {Binnenfeld}, A. and {Blanco-Cuaresma}, S. and {Boch}, T. and {Bombrun}, A. and {Bossini}, D. and {Bouquillon}, S. and {Bragaglia}, A. and {Bramante}, L. and {Breedt}, E. and {Bressan}, A. and {Brouillet}, N. and {Brugaletta}, E. and {Bucciarelli}, B. and {Burlacu}, A. and {Butkevich}, A.~G. and {Buzzi}, R. and {Caffau}, E. and {Cancelliere}, R. and {Cantat-Gaudin}, T. and {Carballo}, R. and {Carlucci}, T. and {Carnerero}, M.~I. and {Carrasco}, J.~M. and {Casamiquela}, L. and {Castellani}, M. and {Castro-Ginard}, A. and {Chaoul}, L. and {Charlot}, P. and {Chemin}, L. and {Chiaramida}, V. and {Chiavassa}, A. and {Chornay}, N. and {Comoretto}, G. and {Contursi}, G. and {Cooper}, W.~J. and {Cornez}, T. and {Cowell}, S. and {Crifo}, F. and {Cropper}, M. and {Crosta}, M. and {Crowley}, C. and {Dafonte}, C. and {Dapergolas}, A. and {David}, P. and {de Laverny}, P.},
  title = {{Gaia Data Release 3. A golden sample of astrophysical parameters}},
  journal = {\aap},
  year = {2023c},
  month = jun,
  volume = {674},
  eid = {A39},
  pages = {A39},
  doi = {10.1051/0004-6361/202243800},
  archiveprefix = {arXiv},
  eprint = {2206.05870},
  primaryclass = {astro-ph.SR},
  adsurl = {https://ui.adsabs.harvard.edu/abs/2023A&A...674A..39G},
}

@article{BB19,
  author = {{Bennett}, Morgan and {Bovy}, Jo},
  title = {{Vertical waves in the solar neighbourhood in Gaia DR2}},
  journal = {\mnras},
  year = {2019},
  month = jan,
  volume = {482},
  number = {1},
  pages = {1417-1425},
  doi = {10.1093/mnras/sty2813},
  archiveprefix = {arXiv},
  eprint = {1809.03507},
  primaryclass = {astro-ph.GA},
  adsurl = {https://ui.adsabs.harvard.edu/abs/2019MNRAS.482.1417B},
}

@article{Gordon23,
  author = {{Gordon}, David and {de Witt}, Aletha and {Jacobs}, Christopher S.},
  title = {{Position and Proper Motion of Sagittarius A* in the ICRF3 Frame from VLBI Absolute Astrometry}},
  journal = {\aj},
  year = {2023},
  month = feb,
  volume = {165},
  number = {2},
  eid = {49},
  pages = {49},
  doi = {10.3847/1538-3881/aca65b},
  archiveprefix = {arXiv},
  eprint = {2212.00632},
  primaryclass = {astro-ph.GA},
  adsurl = {https://ui.adsabs.harvard.edu/abs/2023AJ....165...49G},
}

@article{McMillan22,
  author = {{McMillan}, Paul J. and {Petersson}, Jonathan and {Tepper-Garcia}, Thor and {Bland-Hawthorn}, Joss and {Antoja}, Teresa and {Chemin}, Laurent and {Figueras}, Francesca and {Khanna}, Shourya and {Kordopatis}, Georges and {Ramos}, Pau and {Romero-G{\'o}mez}, Merce and {Seabroke}, George},
  title = {{The disturbed outer Milky Way disc}},
  journal = {\mnras},
  year = {2022},
  month = nov,
  volume = {516},
  number = {4},
  pages = {4988-5002},
  doi = {10.1093/mnras/stac2571},
  archiveprefix = {arXiv},
  eprint = {2206.04059},
  primaryclass = {astro-ph.GA},
  adsurl = {https://ui.adsabs.harvard.edu/abs/2022MNRAS.516.4988M},
}

@article{Bovy15,
  author = {{Bovy}, Jo and {Bird}, Jonathan C. and {Garc{\'\i}a P{\'e}rez}, Ana E. and {Majewski}, Steven R. and {Nidever}, David L. and {Zasowski}, Gail},
  title = {{The Power Spectrum of the Milky Way: Velocity Fluctuations in the Galactic Disk}},
  journal = {\apj},
  year = {2015},
  month = feb,
  volume = {800},
  number = {2},
  eid = {83},
  pages = {83},
  doi = {10.1088/0004-637X/800/2/83},
  archiveprefix = {arXiv},
  eprint = {1410.8135},
  primaryclass = {astro-ph.GA},
  adsurl = {https://ui.adsabs.harvard.edu/abs/2015ApJ...800...83B},
}

@article{Jerabkova21,
  author = {{Jerabkova}, Tereza and {Boffin}, Henri M.~J. and {Beccari}, Giacomo and {de Marchi}, Guido and {de Bruijne}, Jos H.~J. and {Prusti}, Timo},
  title = {{The 800 pc long tidal tails of the Hyades star cluster. Possible discovery of candidate epicyclic overdensities from an open star cluster}},
  journal = {\aap},
  year = {2021},
  month = mar,
  volume = {647},
  eid = {A137},
  pages = {A137},
  doi = {10.1051/0004-6361/202039949},
  archiveprefix = {arXiv},
  eprint = {2103.12080},
  primaryclass = {astro-ph.GA},
  adsurl = {https://ui.adsabs.harvard.edu/abs/2021A&A...647A.137J},
}

@article{Poggio21,
  author = {{Poggio}, E. and {Drimmel}, R. and {Cantat-Gaudin}, T. and {Ramos}, P. and {Ripepi}, V. and {Zari}, E. and {Andrae}, R. and {Blomme}, R. and {Chemin}, L. and {Clementini}, G. and {Figueras}, F. and {Fouesneau}, M. and {Fr{\'e}mat}, Y. and {Lobel}, A. and {Marshall}, D.~J. and {Muraveva}, T. and {Romero-G{\'o}mez}, M.},
  title = {{Galactic spiral structure revealed by Gaia EDR3}},
  journal = {\aap},
  year = {2021},
  month = jul,
  volume = {651},
  eid = {A104},
  pages = {A104},
  doi = {10.1051/0004-6361/202140687},
  archiveprefix = {arXiv},
  eprint = {2103.01970},
  primaryclass = {astro-ph.GA},
  adsurl = {https://ui.adsabs.harvard.edu/abs/2021A&A...651A.104P},
}

@article{Hou21,
  author = {{Hou}, L.~G.},
  title = {{The spiral structure in the Solar neighbourhood}},
  journal = {Frontiers in Astronomy and Space Sciences},
  year = {2021},
  month = aug,
  volume = {8},
  eid = {103},
  pages = {103},
  doi = {10.3389/fspas.2021.671670},
  archiveprefix = {arXiv},
  eprint = {2110.04446},
  primaryclass = {astro-ph.GA},
  adsurl = {https://ui.adsabs.harvard.edu/abs/2021FrASS...8..103H},
}

@article{Zhu26,
  author = {{Zhu}, Ling and {Cai}, Runsheng and {Kang}, Xi and {Xue}, Xiang-Xiang and {Yang}, Chengqun and {Zhang}, Lan and {Mao}, Shude and {Liu}, Chao},
  title = {{A vertically orientated dark matter halo marks a flip of the Galactic disc}},
  journal = {\aap},
  year = {2026},
  month = feb,
  volume = {706},
  eid = {A193},
  pages = {A193},
  doi = {10.1051/0004-6361/202557613},
  archiveprefix = {arXiv},
  eprint = {2510.08684},
  primaryclass = {astro-ph.GA},
  adsurl = {https://ui.adsabs.harvard.edu/abs/2026A&A...706A.193Z},
}

@article{Dillamore26,
  author = {{Dillamore}, Adam M. and {Sanders}, Jason L.},
  title = {{Geometry of the Milky Way's dark matter from dynamical models of the tilted stellar halo}},
  journal = {\mnras},
  year = {2026},
  month = mar,
  volume = {546},
  number = {4},
  eid = {stag226},
  pages = {stag226},
  doi = {10.1093/mnras/stag226},
  archiveprefix = {arXiv},
  eprint = {2510.00095},
  primaryclass = {astro-ph.GA},
  adsurl = {https://ui.adsabs.harvard.edu/abs/2026MNRAS.546ag226D},
}

@article{Shu16,
  author = {{Shu}, Frank H.},
  title = {{Six Decades of Spiral Density Wave Theory}},
  journal = {\araa},
  year = {2016},
  month = sep,
  volume = {54},
  pages = {667-724},
  doi = {10.1146/annurev-astro-081915-023426},
  adsurl = {https://ui.adsabs.harvard.edu/abs/2016ARA&A..54..667S},
}

@article{Gravity19,
  author = {{GRAVITY Collaboration} and {Abuter}, R. and {Amorim}, A. and {Baub{\"o}ck}, M. and {Berger}, J.~P. and {Bonnet}, H. and {Brandner}, W. and {Cl{\'e}net}, Y. and {Coud{\'e} Du Foresto}, V. and {de Zeeuw}, P.~T. and {Dexter}, J. and {Duvert}, G. and {Eckart}, A. and {Eisenhauer}, F. and {F{\"o}rster Schreiber}, N.~M. and {Garcia}, P. and {Gao}, F. and {Gendron}, E. and {Genzel}, R. and {Gerhard}, O. and {Gillessen}, S. and {Habibi}, M. and {Haubois}, X. and {Henning}, T. and {Hippler}, S. and {Horrobin}, M. and {Jim{\'e}nez-Rosales}, A. and {Jocou}, L. and {Kervella}, P. and {Lacour}, S. and {Lapeyr{\`e}re}, V. and {Le Bouquin}, J.-B. and {L{\'e}na}, P. and {Ott}, T. and {Paumard}, T. and {Perraut}, K. and {Perrin}, G. and {Pfuhl}, O. and {Rabien}, S. and {Rodriguez Coira}, G. and {Rousset}, G. and {Scheithauer}, S. and {Sternberg}, A. and {Straub}, O. and {Straubmeier}, C. and {Sturm}, E. and {Tacconi}, L.~J. and {Vincent}, F. and {von Fellenberg}, S. and {Waisberg}, I. and {Widmann}, F. and {Wieprecht}, E. and {Wiezorrek}, E. and {Woillez}, J. and {Yazici}, S.},
  title = {{A geometric distance measurement to the Galactic center black hole with 0.3\% uncertainty}},
  journal = {\aap},
  year = {2019},
  month = may,
  volume = {625},
  eid = {L10},
  pages = {L10},
  doi = {10.1051/0004-6361/201935656},
  archiveprefix = {arXiv},
  eprint = {1904.05721},
  primaryclass = {astro-ph.GA},
  adsurl = {https://ui.adsabs.harvard.edu/abs/2019A&A...625L..10G},
}

@article{Gravity22,
  author = {{GRAVITY Collaboration} and {Abuter}, R. and {Aimar}, N. and {Amorim}, A. and {Ball}, J. and {Baub{\"o}ck}, M. and {Berger}, J.~P. and {Bonnet}, H. and {Bourdarot}, G. and {Brandner}, W. and {Cardoso}, V. and {Cl{\'e}net}, Y. and {Dallilar}, Y. and {Davies}, R. and {de Zeeuw}, P.~T. and {Dexter}, J. and {Drescher}, A. and {Eisenhauer}, F. and {F{\"o}rster Schreiber}, N.~M. and {Foschi}, A. and {Garcia}, P. and {Gao}, F. and {Gendron}, E. and {Genzel}, R. and {Gillessen}, S. and {Habibi}, M. and {Haubois}, X. and {Hei{\ss}el}, G. and {Henning}, T. and {Hippler}, S. and {Horrobin}, M. and {Jochum}, L. and {Jocou}, L. and {Kaufer}, A. and {Kervella}, P. and {Lacour}, S. and {Lapeyr{\`e}re}, V. and {Le Bouquin}, J.-B. and {L{\'e}na}, P. and {Lutz}, D. and {Ott}, T. and {Paumard}, T. and {Perraut}, K. and {Perrin}, G. and {Pfuhl}, O. and {Rabien}, S. and {Shangguan}, J. and {Shimizu}, T. and {Scheithauer}, S. and {Stadler}, J. and {Stephens}, A.~W. and {Straub}, O. and {Straubmeier}, C. and {Sturm}, E. and {Tacconi}, L.~J. and {Tristram}, K.~R.~W. and {Vincent}, F. and {von Fellenberg}, S. and {Widmann}, F. and {Wieprecht}, E. and {Wiezorrek}, E. and {Woillez}, J. and {Yazici}, S. and {Young}, A.},
  title = {{Mass distribution in the Galactic Center based on interferometric astrometry of multiple stellar orbits}},
  journal = {\aap},
  year = {2022},
  month = jan,
  volume = {657},
  eid = {L12},
  pages = {L12},
  doi = {10.1051/0004-6361/202142465},
  archiveprefix = {arXiv},
  eprint = {2112.07478},
  primaryclass = {astro-ph.GA},
  adsurl = {https://ui.adsabs.harvard.edu/abs/2022A&A...657L..12G},
}

@book{Press07,
  author = {{Press}, William H. and {Teukolsky}, Saul A. and {Vetterling}, William T. and {Flannery}, Brian P.},
  title = {{Numerical Recipes: The Art of Scientific Computing}},
  year = {2007},
  adsurl = {https://ui.adsabs.harvard.edu/abs/2007nras.book.....P},
}

@article{Bovy19,
  author = {{Bovy}, Jo and {Leung}, Henry W. and {Hunt}, Jason A.~S. and {Mackereth}, J. Ted and {Garc{\'\i}a-Hern{\'a}ndez}, Domingo A. and {Roman-Lopes}, Alexandre},
  title = {{Life in the fast lane: a direct view of the dynamics, formation, and evolution of the Milky Way's bar}},
  journal = {\mnras},
  year = {2019},
  month = dec,
  volume = {490},
  number = {4},
  pages = {4740-4747},
  doi = {10.1093/mnras/stz2891},
  archiveprefix = {arXiv},
  eprint = {1905.11404},
  primaryclass = {astro-ph.GA},
  adsurl = {https://ui.adsabs.harvard.edu/abs/2019MNRAS.490.4740B},
}

@article{Bernet24,
  author = {{Bernet}, M. and {Ramos}, P. and {Antoja}, T. and {Monari}, G. and {Famaey}, B.},
  title = {{Radial and azimuthal gradients of the moving groups in Gaia DR3: The slow and fast bar degeneracy problem}},
  journal = {\aap},
  year = {2024},
  month = jun,
  volume = {686},
  eid = {A92},
  pages = {A92},
  doi = {10.1051/0004-6361/202348410},
  archiveprefix = {arXiv},
  eprint = {2403.02393},
  primaryclass = {astro-ph.GA},
  adsurl = {https://ui.adsabs.harvard.edu/abs/2024A&A...686A..92B},
}

@article{Bovy12,
  author = {{Bovy}, Jo and {Allende Prieto}, Carlos and {Beers}, Timothy C. and {Bizyaev}, Dmitry and {da Costa}, Luiz N. and {Cunha}, Katia and {Ebelke}, Garrett L. and {Eisenstein}, Daniel J. and {Frinchaboy}, Peter M. and {Garc{\'\i}a P{\'e}rez}, Ana Elia and {Girardi}, L{\'e}o and {Hearty}, Fred R. and {Hogg}, David W. and {Holtzman}, Jon and {Maia}, Marcio A.~G. and {Majewski}, Steven R. and {Malanushenko}, Elena and {Malanushenko}, Viktor and {M{\'e}sz{\'a}ros}, Szabolcs and {Nidever}, David L. and {O'Connell}, Robert W. and {O'Donnell}, Christine and {Oravetz}, Audrey and {Pan}, Kaike and {Rocha-Pinto}, Helio J. and {Schiavon}, Ricardo P. and {Schneider}, Donald P. and {Schultheis}, Mathias and {Skrutskie}, Michael and {Smith}, Verne V. and {Weinberg}, David H. and {Wilson}, John C. and {Zasowski}, Gail},
  title = {{The Milky Way's Circular-velocity Curve between 4 and 14 kpc from APOGEE data}},
  journal = {\apj},
  year = {2012},
  month = nov,
  volume = {759},
  number = {2},
  eid = {131},
  pages = {131},
  doi = {10.1088/0004-637X/759/2/131},
  archiveprefix = {arXiv},
  eprint = {1209.0759},
  primaryclass = {astro-ph.GA},
  adsurl = {https://ui.adsabs.harvard.edu/abs/2012ApJ...759..131B},
}

@article{Reid19,
  author = {{Reid}, M.~J. and {Menten}, K.~M. and {Brunthaler}, A. and {Zheng}, X.~W. and {Dame}, T.~M. and {Xu}, Y. and {Li}, J. and {Sakai}, N. and {Wu}, Y. and {Immer}, K. and {Zhang}, B. and {Sanna}, A. and {Moscadelli}, L. and {Rygl}, K.~L.~J. and {Bartkiewicz}, A. and {Hu}, B. and {Quiroga-Nu{\~n}ez}, L.~H. and {van Langevelde}, H.~J.},
  title = {{Trigonometric Parallaxes of High-mass Star-forming Regions: Our View of the Milky Way}},
  journal = {\apj},
  year = {2019},
  month = nov,
  volume = {885},
  number = {2},
  eid = {131},
  pages = {131},
  doi = {10.3847/1538-4357/ab4a11},
  archiveprefix = {arXiv},
  eprint = {1910.03357},
  primaryclass = {astro-ph.GA},
  adsurl = {https://ui.adsabs.harvard.edu/abs/2019ApJ...885..131R},
}

\label{lastpage}

\end{document}